\documentclass[journal]{IEEEtran}
\title{Optimal control of end-user energy storage}
\author{Peter M. van de Ven, Nidhi Hegde, Laurent Massouli\'e and Theodoros Salonidis,~\IEEEmembership{Member,~IEEE}%
\thanks{Peter M. van de Ven and Theodoros Salonidis are with the IBM Thomas J.\ Watson Research Center, Nidhi Hegde is with Technicolor Paris Research Lab and Laurent Massouli\'e is with INRIA, MSR-Inria Joint Centre.}%
\thanks{A preliminary version of this paper has appeared as~\cite{VHMS11}.}
}

\usepackage{amsmath,amssymb,cite,dsfont,graphicx,psfrag,subfigure}
\usepackage[english]{babel}
\usepackage{color}

\newcommand{\Amax}{\bar{A}}
\newcommand{\Bmax}{\bar{B}}
\newcommand{\Bu}{\mathcal{B}}
\newcommand{\Pmax}{\bar{P}}
\newcommand{\Pmin}{p_{\rm min}}

\newcommand{\Dmax}{\bar{D}}

\newcommand{\bfx}{\boldsymbol x}
\newcommand{\bfy}{\boldsymbol y}

\newcommand{\sright}{\sigma_{\bfx}^+}
\newcommand{\sleft}{\sigma_{\bfx}^-}
\newcommand{\Uright}{U_{\bfx}^+(b)}
\newcommand{\Uleft}{U_{\bfx}^-(b)}

\newcommand\1{\mathbf{1}}
\newcommand{\indi}[1]{\1_{\{#1\}}}
\newcommand{\expect}[1]{{\mathbb E}\{#1\}}

\newtheorem{corollary}{Corollary}
\newtheorem{lemma}{Lemma}
\newtheorem{proposition}{Proposition}
\newtheorem{theorem}{Theorem}
\newtheorem{example}{Example}
\newtheorem{remark}{Remark}

\graphicspath{{Images/}}

\psfrag{A1}{$A_1(t)$}
\psfrag{A2}{$A_2(t)$}
\psfrag{A3}{$A_3(t)$}
\psfrag{A}{$A_3(t)$}
\psfrag{Amax}{$\Amax$}
\psfrag{Bbar}{$\Bmax {\rm ~(kWh)}$}
\psfrag{beta}{$\beta {\rm ~(kWh)}$}
\psfrag{beta1}{$\beta_1$}
\psfrag{beta2}{$\beta_2$}
\psfrag{beta3}{$\beta_3$}
\psfrag{beta_avr}{$\beta_{\bfx}$}
\psfrag{betap}{$\beta(p)$}
\psfrag{beta_min}{$\beta_{\bfx}^-$}
\psfrag{beta_plus}{$\beta_{\bfx}^+$}
\psfrag{bopt}{$\beta_{\bfx}^*$}
\psfrag{Bmax_min_d}{$\Bmax - d(\bfx)$}
\psfrag{smallb}{$b$}
\psfrag{Bt}{$B(t)$}
\psfrag{BpminA}{$\beta(p) - \bar{A}$}
\psfrag{Btplus}{$B^*(t+1)$}
\psfrag{BplusD}{$\beta(p) + D(t)$}
\psfrag{Deltat}{$\Delta(t)$}
\psfrag{Dt}{$D(t)$}
\psfrag{mind}{$-D(t)$}
\psfrag{Uplus}{$U_{\bfx}^+(b)$}
\psfrag{Umin}{$U_{\bfx}^-(b)$}

\begin{document}
\maketitle

%\footnotetext[1]{Department of Mathematics \& Computer Science,
%Eindhoven University of Technology
%P.O. Box 513, 5600 MB Eindhoven, The Netherlands}
%\footnotetext[2]{EURANDOM, P.O. Box 513, 5600 MB Eindhoven, The Netherlands}
%\footnotetext[3]{Technicolor Paris Research Lab, 75015 Paris, France}

\begin{abstract}
An increasing number of retail energy markets show price fluctuations, providing users with the opportunity to buy energy at lower than average prices. We propose to temporarily store this inexpensive energy in a battery, and use it to satisfy demand when energy prices are high, thus allowing users to exploit the price variations without having to shift their demand to the low-price periods. We study the battery control policy that yields the best performance, i.e., minimizes the total discounted costs. The optimal policy is shown to have a threshold structure, and we derive these thresholds in a few special cases. The cost savings obtained from energy storage are demonstrated through extensive numerical experiments, and we offer various directions for future research.
\end{abstract}


\section{Introduction}\label{sec:intro}

%An increasing number of electricity markets are replacing the traditional fixed retail energy rates by prices that fluctuate during the day, to better reflect the time-varying costs for generating  energy. This provides an incentive for customers to shift their demand from peak-hours to off-peak hours, but in practice the demand elasticity of retail customers is insufficient to successfully exploit the price variations. By temporarily storing inexpensive energy in a battery for later use, consumers can benefit from the fluctuating price without having to adjust their demand pattern. We are interested in the cost-minimizing policy for charging and discharging the battery.

Wholesale energy prices exhibit significant fluctuations during each day due to variations in demand and generator capacity. End users are traditionally not exposed to these fluctuations but pay a fixed retail energy price%, as shown in Figure~\ref{fig:price2}
. Economists have long argued to remove the fixed retail prices in favor of prices that change during the day. Such {\em dynamic pricing} would better reflect the prices on the wholesale market and has been predicted to lead to lower demand peaks and to a lower and less volatile wholesale price~\cite{Borenstein05}. Implementations of dynamic pricing have been enabled by recent developments in smart-grid technology such as smart meters.

An example of an increasingly popular dynamic pricing scheme
is time-of-use pricing. %(Figure~\ref{fig:price3}).
Such pricing typically provides two
or three price levels (e.g., `off-peak', `mid-peak' and `on-peak') where the relevant level
depends on the time of day. The price levels are determined well in advance and are
only changed once or twice per year. A second example of dynamic
pricing is real-time pricing %(Figure~\ref{fig:price4})
where the retail energy price
changes hourly or half-hourly and are based on the price on the wholesale energy market.
%This approach is more flexible than time-of-use pricing but raises implementation issues such as keeping the users informed about the energy price.
%\begin{figure}[h]
% \begin{center}
% %\subfigure[wholesale price]{\label{fig:price1} \includegraphics[width = 0.4\textwidth]{price1}}\hspace{0.5cm}
% \subfigure[fixed pricing]{\label{fig:price2} \includegraphics[width = 0.4\textwidth]{price2}}
% \subfigure[time-of-use pricing]{\label{fig:price3} \includegraphics[width = 0.4\textwidth]{price3}}\hspace{0.5cm}
% \subfigure[real-time pricing]{\label{fig:price4} \includegraphics[width = 0.4\textwidth]{price4}}
% \end{center}
% \caption{The wholesale energy price (gray) and various approaches to retail pricing (black).}
%\label{fig:price}
%\end{figure}

Dynamic pricing creates an opportunity for users such as households and data centers to exploit the price fluctuations and reduce energy costs. However, doing so would require users to shift their demands to low-price periods, and in practice users only show a minor shift in their demand in response to changes in the energy prices~\cite{AE08,Allcott11,BGN04,Lijesen07}.
%For example, a current pilot project with real-time pricing shows that users only marginally decrease their activity when prices are high.
A possible solution is to equip users with a battery that can be used for energy storage; the battery can be charged when the energy price is low and the stored energy can then be used when the price is high. This allows users to benefit from the energy price variations without having to adjust their consumption. Energy can be stored both by a dedicated battery, or by existing storage such as the battery pack of an electric car~\cite{PWA10} (residential users) or a backup power supply~\cite{UUNS11} (data centers). %In the past such setup was not economically viable due to the high cost of batteries, but current developments have brought storage applications within reach.

In this paper, we address the problem of organizing energy storage purchases to minimize long-term energy costs under variable demands and prices. This problem involves deciding whether to satisfy demand directly from the grid or from the battery, as well as up to what level to charge or discharge the battery.
%This decision is complicated by the fact that the battery may not be fully efficient and energy is lost during the charging and discharging of the battery.
The resulting optimization problem is complicated by the stochastic nature of price and demand and due to the fact that we aim to minimize the long-term costs. We model the problem as a Markov Decision Process and show that there exists a two-threshold stationary cost-minimizing policy. When the battery level is below the lower threshold, the battery is charged up to it, and the battery is discharged when above the upper threshold. By comparing the costs incurred under this policy with the cost of satisfying all demand directly from the grid, we can show that energy storage may lead to significant cost savings.

%To the best of our knowledge, previous work on energy storage is limited and does not propose optimal control solutions subject to stochastic price and demand fluctuations.
Residential energy storage has been studied for the case of {\em arbitrage}, i.e., buying energy when it is inexpensive, and selling it later to the grid for a higher price~\cite{GJM99}. This problem has been studied assuming that prices are known in advance in a finite horizon setting. These assumptions allow deterministic optimization problem formulations which can be solved using linear programming techniques~\cite{AD09,HCB10}. However, such approach does not take into account
the stochasticity in prices and demands and does not allow for long-term cost optimization. % over infinite horizon.
%Our approach can be readily adapted to an arbitrage problem in an infinite-horizon setting, where the behavior of the price process may be stochastic.
%A similar threshold-based optimal policy can be shown to hold in this case.
%We recently became aware of a related paper~\cite{UUNS11} that investigates optimal control of energy storage in the context of data centers. The model in this paper is similar to ours, with the exception that it is assumed in~\cite{UUNS11} that the battery is fully efficient. Our approach is different in that we are interested in the structure of the optimal policy, while in~\cite{UUNS11} a simple heuristic is presented that is suboptimal but readily implementable.
In~\cite{KHT11} the authors consider the problem of energy storage from the point of view of the grid operator, and propose a threshold policy that is shown to be asymptotically optimal as the size of the storage unit increases.

%We also recently became aware of a parallel work that uses
A model similar to ours was used to investigate control of energy storage
in the context of {\em data centers}~\cite{UUNS11}.
The model in~\cite{UUNS11} assumes that the battery is fully efficient
and the proposed scheduling algorithm is a sub-optimal heuristic,
whose gap from optimality increases as storage size decreases. In~\cite{GDFW11} this approach is extended to multiple data centers, each with different time-varying prices.
In contrast to~\cite{UUNS11} and~\cite{GDFW11}, our model incorporates battery inefficiencies
and we investigate the optimal scheduling policy.

A related problem is optimal control of energy storage for renewable energy. The two-price case is considered in~\cite{HD11}, while the more general case is discussed in~\cite{ZSSS11} and~\cite{HD12}. The case without transmission costs in~\cite{ZSSS11} is closely related to our setting, and a similar two-threshold policy is shown to be optimal. In recent work~\cite{HD12}, the authors consider a similar model to ours to address storage control for renewable energy generation for general prices, and show that the average cost optimal policy has a similar threshold structure. The authors of~\cite{ZSSS11} and~\cite{HD12} use a finite-horizon and infinite horizon average criterion, respectively. The battery and price models differ slightly, and the work in~\cite{HD12} accounts for dissipation losses, but does not allow state-dependent charging constraints. In contrast to~\cite{HD12}, our approach incorporates periodicity, which allows us to model daily fluctuations in price and demand.

Our model is closely related to periodic-review, single-item inventory models, and the optimal policy mirrors the optimality of the base-stock policy for the backlog case. However, in our case the demand is known before the purchasing decision, and we require that all demand is met in the time slot that it arises. In addition, the state description is continuous, and the battery inefficiency fundamentally changes the structure and analysis of the optimal policy.

The remainder of the paper is structured as follows. In Section~\ref{sec:model} we introduce the model and describe the decision variables. In Section~\ref{sec:structure} we demonstrate the optimality of a threshold policy, and in Section~\ref{sec:thresholds} we derive some properties of the thresholds. Section~\ref{sec:example} discusses various numerical examples and Sections~\ref{sec:outlook} and \ref{sec:conclusions} describe some directions for future research and provides concluding remarks, respectively.

\section{Model}\label{sec:model}

Consider a user with certain energy requirements and a battery that can be used for energy storage. Time is slotted, and we denote by $B(t)$ the battery level (state of charge) in kWh at time $t$, $t = 0,1,\dots$. Let $\Bmax$ represent the maximum battery level, and $\Bu = [0,\Bmax]$ the range of all possible battery levels, so
\begin{equation}\label{eqn:constr1}
B(t) \in \Bu.
\end{equation}
In each time slot $t$ some demand $D(t)$ arises, and we may purchase energy at a price of $P(t)$ per unit. The demand has some compact support $D(t) \in \mathcal{D}$, as does the price $P(t) \in \mathcal{P}$. Both are bounded, and we denote by $\Dmax$ and $\Pmax$ the maximum demand and price, respectively, so $\mathcal{D} \subseteq [0,\Dmax]$ and $\mathcal{P} \subseteq [\Pmin,\Pmax]$, with $\Pmin$ the minimum price. Finally, we denote by $\mathcal{M}$ the set of modulating states, that influence the price and demand transitions. For example the time of day or the season.

Denote by $\Omega = \mathcal{D} \times \mathcal{P} \times \mathcal{M}$ the set of possible realizations of demand, price and modulating state, and for any $\bfx \in \Omega$, denote by $d(\bfx)$ and $p(\bfx)$ the corresponding price and demand, respectively. Demand and price may be correlated, and we denote by $f_{\bfx}(\bfy)$ the probability density function of moving from state $\bfx$ to state $\bfy$ in the next slot, for any $\bfx,\bfy \in \Omega$.

The battery may not be completely efficient, and its performance is affected by the charging efficiency  $\eta_c \in (0,1]$ and discharging efficiency $\eta_d \in (0,1]$. Energy purchased to charge the battery is reduced by a factor $\eta_c$, and only a fraction $\eta_d$ of the discharged energy is converted into electricity.
This model is general and encompasses for example batteries of electric vehicles, uninterrupted power supplies of data centers and batteries dedicated to end user storage.  The model is not intended to capture all the subtleties of battery behavior in each of these applications but rather the essential tradeoffs and phenomena that arise in practice.
%This battery model is very general, and encompasses for example battery electric vehicles, uninterrupted power supplies used in data centers, and batteries dedicated to end-user storage. Naturally we are unable to model all the subtleties of battery behavior, but our model captures some of the essential tradeoffs and phenomena that arise in practice.
In Section~\ref{sec:outlook} we discuss various model refinements that can be made without fundamentally altering the results or the derivations.

\begin{figure}[h]
    \begin{center}
        \includegraphics[width=0.85\linewidth]{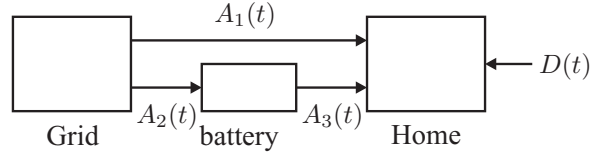}
    \end{center}
    \caption{A graphical representation of the model.}
    \label{fig:house}
\end{figure}

In addition to satisfying the demand from the battery, we also allow demand to be met directly from the grid, bypassing the battery. Let $A_1(t)$ denote the amount of energy purchased directly from the grid in slot $t$, $A_2(t)$ the amount of energy bought to charge the battery, and $A_3(t)$ the energy discharged from the battery used towards satisfying demand, see Figure~\ref{fig:house}.

We assume
\begin{equation}\label{eqn:constr_pos}
A_1(t), A_2(t), A_3(t) \ge 0,
\end{equation}
and require that all demand must be met, i.e.,
\begin{equation}\label{eqn:constr2}
D(t) = A_1(t) + \eta_d A_3(t).
\end{equation}
The battery has state-dependent charging constraints $\Amax_c(b)$ and $\Amax_d(b)$, so the amount of energy that can be charged to and discharged from the battery is bounded as
\begin{equation}\label{eqn:constr_rate}
A_2(t) \le \Amax_c(b), \quad A_3(t) \le \Amax_d(b).
\end{equation}

The battery level of the battery evolves according to
\begin{equation}\label{eqn:constr3}
B(t+1) = B(t) + \eta_c A_2(t) - A_3(t),
\end{equation}
and the energy costs in slot~$t$ are given by
\begin{equation}\label{eqn:energy_costs}
g(t) = (A_1(t) + A_2(t)) P(t).
\end{equation}
We are interested in the total discounted costs $\sum_{t = 0}^\infty g(t) \alpha^t$, with $0 < \alpha < 1$ the discount factor that represents value reduction over time. The reason for considering discounted rather than average costs is to emphasize early decisions and costs, in order to emulate the effect of reduced battery efficiency over time. Note that the total discounted costs are finite, since the per-slot costs are bounded. Our goal is now to choose in each slot the $A_1(t)$, $A_2(t)$ and $A_3(t)$ that solve the following optimization problem:
\begin{align}
\nonumber &\min \expect{\sum_{t = 0}^\infty g(t) \alpha^t}\\
&{\rm subject~to:}~\eqref{eqn:constr1},\eqref{eqn:constr_pos},\eqref{eqn:constr2},\eqref{eqn:constr_rate},\eqref{eqn:constr3} \label{eqn:min_problem}
\end{align}
%Note that the costs~\eqref{eqn:energy_costs} are limited to the costs for purchasing energy. Although most results in this paper can be extended to include for example costs for replacing the battery, this is not the focus of this paper.

%The infinite-horizon problem~\eqref{eqn:min_problem} can be recognized as a stochastic optimization problem,
The infinite horizon problem belongs to a class of stochastic optimization problems, which
in general are difficult to solve. A first step towards dealing with~\eqref{eqn:min_problem} is to see that it is never optimal to charge and discharge the battery in the same slot.
%\begin{proposition}\label{pro:simplify}
%Let $t \ge 0$ and denote by $A^*_1(t)$,$A^*_2(t)$,$A^*_3(t)$ the choices that solve~\eqref{eqn:min_problem}. Then
%\begin{equation}
%A^*_2(t)A^*_3(t) = 0.
%\end{equation}
%\end{proposition}
%
%The proof of Proposition~\ref{pro:simplify}, can be found in the appendix.
This is intuitively clear, because charging and discharging the battery simultaneously corresponds to routing $\min\{A_2(t),\eta_d A_3(t)\}$ energy from the grid to the user, through the battery. Because of the battery inefficiency it is beneficial to instead circumvent the battery, and satisfy the demand directly from the grid.

It turns out that this observation significantly simplifies the optimization problem, by reducing the number of decision variables.
Specifically, %denote by $\Delta(t)$ the difference in buffer level between slot~$t$ and~$t+1$, i.e.,
%\begin{equation*}
%B(t+1) = B(t) + \Delta(t).
%\end{equation*}
%Then,
in view of the restriction on simultaneous charging and discharging:
\begin{align*}
A_1(t) &= D(t) + (B(t+1) - B(t)) \eta_d \indi{B(t+1) < B(t)},\\
A_2(t) &= (B(t+1) - B(t)) \eta_c^{-1} \indi{B(t+1) > B(t)},\\
A_3(t) &= -(B(t+1) - B(t)) \eta_d \indi{B(t+1) < B(t)}.
\end{align*}
Thus, the choice for $B(t+1)$ fixes $A_1(t)$, $A_2(t)$ and $A_3(t)$, and~\eqref{eqn:min_problem} reduces to a single-variable decision problem. The per-slot costs may be rewritten in terms of $B(t+1)$ as
\begin{align*}
g(t) ={}& (D(t) + (B(t+1) - B(t))^+ \eta_c^{-1}\\
& + (B(t+1) - B(t))^- \eta_d) P(t),
\end{align*}
with $(x)^+ = \max\{x,0\}$ and $(x)^- = -\max\{-x,0\}$. Note that~\eqref{eqn:constr_rate} can be expressed in $B(t+1)$ as
\begin{equation*}
-\Amax_d(B(t)) \le B(t+1) - B(t) \le \bar{A}_c(B(t)) \eta_c.
\end{equation*}

\section{The structure of the optimal policy}\label{sec:structure}
In this section we discuss how to choose in each slot the value for $\beta = B(t+1)$ that minimizes the total discounted costs. To this end, we rewrite our model as a Markov decision process. We denote by $J_{\bfx}(b)$ the minimal total discounted costs for a battery differential $\delta$, starting from state $\bfx \in \Omega$, and battery level $b \in \Bu$. Let $\gamma_{\bfx}(\delta) = (d(\bfx) + (\delta)^+ \eta_c^{-1} + (\delta)^- \eta_d) p(\bfx)$ denote the immediate costs, $G_{\bfx}(\beta) = \int\limits_{\bfy \in \Omega} f_{\bfx}(\bfy) J_{\bfy}(\beta){\rm d}\bfy$ and define
\begin{equation*}
H_{\bfx}(\beta,b) = \gamma_{\bfx}(\beta - b) + \alpha G_{\bfx}(\beta),
\end{equation*}
the total discounted costs given battery level $b$, state $\bfx$ and action $\beta$. Then the cost function satisfies the Bellman equation
\begin{equation}\label{eqn:bellman}
J_{\bfx}(b) = \inf_{\beta \in U_{\bfx}(b)}H_{\bfx}(\beta,b),
\end{equation}
with $U_{\bfx}(b)$ the control set that contains all feasible decisions for $B(t+1)$. This set may be written as
\begin{equation*}
U_{\bfx}(b)  = [\Uleft,\Uright],
\end{equation*}
where $\Uleft = \max\{0,b - d(\bfx),b-\Amax_d(b)\}$ and $\Uright = \min\{\Bmax, b + \Amax_c(b)\}$.

It is readily verified that the infimum in~\eqref{eqn:bellman} can be attained by a stationary optimal policy, see, e.g.,~\cite[Proposition 4.4]{BS96}. We shall demonstrate that this optimal policy specifies for each state $\bfx \in \Omega$ two battery thresholds $\beta_{\bfx}^-,\beta_{\bfx}^+ \in \Bu$, $\beta_{\bfx}^- \le \beta_{\bfx}^+$ such that the cost-minimizing choice for the battery level $\beta_{\bfx}^*(b)$ is given by
\begin{equation}\label{eqn:optimal_policy}
\beta_{\bfx}^*(b) = \left\{
\begin{array}{ll}
\min\{\beta_{\bfx}^-, \Uright\}, & b \le \beta_{\bfx}^-,\\
\max\{\beta_{\bfx}^+, \Uleft\}, & b \ge \beta_{\bfx}^+,\\
b,  &   {\rm otherwise}.
\end{array}
\right.
\end{equation}
So if $b \le \beta_{\bfx}^-$, then the optimal policy is to charge the battery as close to $\beta_{\bfx}^-$ as the control set allows. If $b \ge \beta_{\bfx}^+$, then the battery should be discharged down to $\beta_{\bfx}^+$ within the boundaries of the control set. In case $\beta_{\bfx}^- < b < \beta_{\bfx}^+$, it is optimal to neither charge nor discharge the battery. So if the battery level is sufficiently low, the battery is charged, and all demand in satisfied directly from the grid, while for high battery level, demand is (partially) satisfied from the battery. When the battery level is between both thresholds, the battery is neither charged nor discharged, and all demand is met from the grid.

In order to show that~\eqref{eqn:optimal_policy} indeed describes the structure of the optimal policy, we require the following lemma.

\begin{lemma}\label{lem:cost_properties}
The cost function $J_{\bfx}(b)$ is convex and non-increasing in $b$.
\end{lemma}

The proof of Lemma~\ref{lem:cost_properties} relies on the fact that the total discounted costs can be viewed as the limit of finite-horizon discounted costs, which may be shown to possess these properties by induction on the horizon. The proof can be found in the appendix, along with the other proofs.

%, and can be found in Section~\ref{sec:proof_lemma}.

We are now in position to state and prove our main result.
\begin{theorem}\label{thm:optimal}
The policy that solves the minimization problem~\eqref{eqn:min_problem} is of the form $\beta = \beta_{\bfx}^*$  as in~\eqref{eqn:optimal_policy}.
\end{theorem}

%The proof of Theorem~\ref{thm:optimal} can be found in Section~\ref{sec:proof_theorem}.
By Lemma~\ref{lem:cost_properties} we know that the right-hand side of the Bellman equation~\eqref{eqn:bellman} in fact defines a convex optimization problem, the solution of which can be found by finding the $\beta \in U_{\bfx}(b)$ for the which the subdifferential $\partial H_{\bfx}(\beta,b)$ has the proper shape. The proof of Theorem~\ref{thm:optimal} then relies on close inspection of this subdifferential.

In case the battery is fully efficient, the charging threshold and discharging threshold are identical, as is shown in the next corollary.

\begin{corollary}\label{col:efficient}
Let $\eta_c = \eta_d = 1$, then the optimal policy is as in~\eqref{eqn:optimal_policy}, with $\beta_{\bfx}^- = \beta_{\bfx}^+$.
\end{corollary}

In Section~\ref{sec:outlook} we present several model extensions under which Theorem~\ref{thm:optimal} remains valid.

%\begin{remark}
%Theorem~\ref{thm:optimal} can be readily extended to incorporate energy arbitrage, by allowing for energy to be sold to the grid. Specifically, we allow negative $A_1(t)$, so it is possible to take more energy from the battery than needed to satisfy the demand ($\eta_d A_3(t) > D(t)$), and the remaining energy will be sold back to the grid $A_1(t) = D(t) - \eta_d A_3(t)$. Moreover, our model may be adapted to allow for energy generation by the user (e.g., solar or wind energy). This can be done by removing the constraint $D(t) \ge 0$, and by redefining $D(t)$ as the difference between the demand and the amount of locally generated energy in slot~$t$. When $D(t) < 0$, this means that more energy was generated than can be used, and the remaining energy can be stored or sold to the grid.
%\end{remark}

\begin{figure}[h]
    \begin{center}
        \includegraphics[width=1\linewidth]{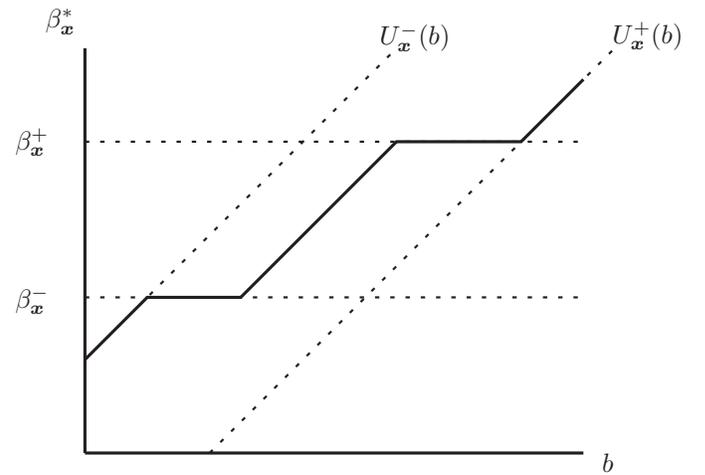}
    \end{center}
    \caption{The structure of the optimal policy $\beta_{\bfx}^*$ as a function of $b$.}
    \label{fig:policy}
\end{figure}

The optimal policy~\eqref{eqn:optimal_policy} is illustrated in Figure~\ref{fig:policy}, which shows $\beta_{\bfx}^*(b)$ plotted against $b$. The diagonal segment in the middle corresponds to $\beta_{\bfx}^*(b) = b$, while the two horizontal lines represents the thresholds $\beta_{\bfx}^-$ and $\beta_{\bfx}^+$. The outer diagonal segments represent the boundaries of the control space $U_{\bfx}(b)$. Both horizontal lines coincide in the scenario with a completely efficient battery ($\eta_d \eta_c = 1$).

%\begin{figure}[h]
% \begin{center}
% \subfigure[$\eta_d \eta_c < 1$]{\label{fig:policy_inefficient} \includegraphics[width = 0.5\textwidth]{policy_inefficient}}\hspace{0.5cm}
% \subfigure[$\eta_d \eta_c = 1$]{\label{fig:policy_efficient} \includegraphics[width = 0.5\textwidth]{policy_efficient}}
%  \end{center}
% \caption{The structure of the optimal policy $\beta_{\bfx}^*$ as a function of $b$.}
%\label{fig:policy}
%\end{figure}

\section{Battery level thresholds}\label{sec:thresholds}

In the previous section we have established the threshold-based structure of the optimal policy~\eqref{eqn:optimal_policy}. Analytically determining the thresholds $\beta_{\bfx}^-$ and $\beta_{\bfx}^+$ is a difficult problem in general, and in this section we present some structural results and results for special cases. For simplicity, we limit ourselves to the case $\eta_d = \eta_c = 1$, and denote $\beta_{\bfx} = \beta_{\bfx}^- = \beta_{\bfx}^+$. Similar properties can be shown to hold for the inefficient case.

We first present sufficient conditions for the thresholds to be equal to either 0 or $\Bmax$.

\begin{proposition}\label{pro:sufficient_conditions}
Let $\bfx \in \Omega$. If for all $b_1,b_2 \in \Bu$, $b_1 < b_2$, $\bfy \in \Omega$
\begin{equation*}
J_{\bfy}(b_1) - J_{\bfy}(b_2) \le (b_2 - b_1) \frac{p(\bfx)}{\alpha},
\end{equation*}
then $\beta_{\bfx} = 0$. If for all $b_1,b_2 \in \Bu$, $b_1 < b_2$, $\bfy \in \Omega$
\begin{equation*}
J_{\bfy}(b_1) - J_{\bfy}(b_2) \ge (b_2 - b_1) \frac{p(\bfx)}{\alpha},
\end{equation*}
then $\beta_{\bfx} = \Bmax$.
\end{proposition}

This result can for example be used to show that $\beta_{\bfx} = 0$ in certain special cases, as is described in the following proposition.

\begin{proposition}\label{pro:max_price}
Let $\bfx \in \Omega$ and denote $p_{\rm min} = \min_{p \in \mathcal{P}} p$, then $\beta_{\bfx} = 0$ if either:\\
(i) $p(\bfx) = \Pmax$;\\
(ii) $p_{\rm min} >0$ and $\alpha < p_{\rm min}/\Pmax.$
\end{proposition}

Proposition~\ref{pro:max_price} states that if the price is very high, or the discount factor is sufficiently low, it is optimal not to charge the battery at all, and to try to satisfy the demand from the battery as much as possible.

In case that the state transitions are i.i.d.\ ($f_{\bfx} \equiv f$) or if the transition probabilities are determined by the price level ($f_{\bfx} \equiv f_{p(\bfx)}$), the thresholds $\beta_{\bfx}$ depend on the price only and are independent from the demand and modulating state. In this case, write $\beta_{\bfx} = \beta_{p(\bfx)}$. We can show that for i.i.d.\ prices, the thresholds are decreasing in the price level.
\begin{proposition}\label{pro:monotone_thresholds}
Assume that the prices and demands are i.i.d.\ across time, then $\beta_{p(\bfx)}$ is decreasing in $p$.
\end{proposition}

The monotonicity observed in Proposition~\ref{pro:monotone_thresholds} is very intuitive, but does not extend to Markovian prices. The reason is that threshold values are partially determined by the evolution of the price: even for a low price it might be beneficial to set a low threshold, if the price in the next slot is even lower. Such an example is presented below. For ease of presentation we use a discrete probability distribution $f$ in this example, noting that similar examples can be constructed using continuous transition probabilities.
\begin{example}\label{exa:example}
Consider an example with four price levels $p_i = i$, $i = 1,\dots,4$. We assume that $\alpha \ge 3/4$, $\Bmax = 1$ and $D \equiv 1$, and we choose the following price transition probabilities:
\begin{equation*}
f_1(1) = f_1(3) = 1/2, \quad f_2(1) = f_3(4) = f_4(2) = 1.
\end{equation*}
The corresponding thresholds are $\beta_1 = \beta_3 = 1$ and $\beta_2 = \beta_4 = 0$, which are clearly not monotone. The derivation of these thresholds is presented in Appendix~\ref{sec:details_example}.
\end{example}

%\begin{proposition}
%{\bf something on $\beta = 0$ when the price or discount factor is low}.
%\end{proposition}
%
%{\bf remark numerical procedures, difficult for continuous, may discretize but curse of dimensionality. Approximate dynamic programming?}

\section{Numerical evaluation}\label{sec:example}

In this section we evaluate the operation of the optimal energy storage management policy~\eqref{eqn:optimal_policy} in residential environments and real-time pricing (RTP) scenarios.
%Our goal is not an exhaustive evaluation of the
%optimal policy but a demonstration of its practical feasibility and the extent of its cost savings under price and demand fluctuations that might arise in real life {(\bf FIXME: we may or may not want to use this sentence)}.
Our goal is to demonstrate the practical feasibility of the optimal policy and the extent
of its cost savings under RTP and demand fluctuations that might arise in real life.
We first describe the price and demand datasets used for the evaluation and then outline
our low complexity implementation of the optimal policy.
We evaluate the cost savings of the optimal policy in scenarios of
individual home storage units and shared energy storage.
%More specifically, we consider residential users in Ontario subject to RTP and compute their electricity
%costs for February 2011 in two scenarios: with an optimally controlled storage unit and without storage.

\subsection{Price and demand datasets}

We emulate residential RTP data with historical hourly
spot prices of the Ontario energy market~\cite{Ontario2010}. Although
Ontario currently does not use RTP, the spot prices provide
a reasonable RTP estimate.
The residential demand data is synthetically generated using
the tool in~\cite{RTIC10}. This tool uses a high-resolution model of
domestic electricity usage based on patterns of home
occupancy and appliance usage, weather conditions and characteristics of all major appliances commonly found in the
domestic environment.

In our approach, the optimal thresholds can be dynamically determined using empirical
distributions of historical price and demand data for each time slot of the day.
The smallest time slot is determined by the coarsest granularity between
price and demand data. In our experiments, we have hourly price data
and 1-minute demand data, therefore we use time slots of one hour.
We also use a training window of one month.
We found that this window size is long enough
to provide adequate characterization of the distribution of each hour of the day,
and short enough to use the optimal thresholds for the prices
and demands that appear in the next window.

For concise presentation, we show results for a representative scenario
where a home of four occupants equipped with a battery receives hourly prices
from the Ontario energy market during January and February 2011.
%The demand realizations (1 minute granularity) for each day
%of January and February are generated using the tool in~\cite{RTIC10}.

\begin{figure}[h]
    \begin{center}
        \subfigure[January]{\label{fig:jan} \includegraphics[width = 0.23\textwidth]{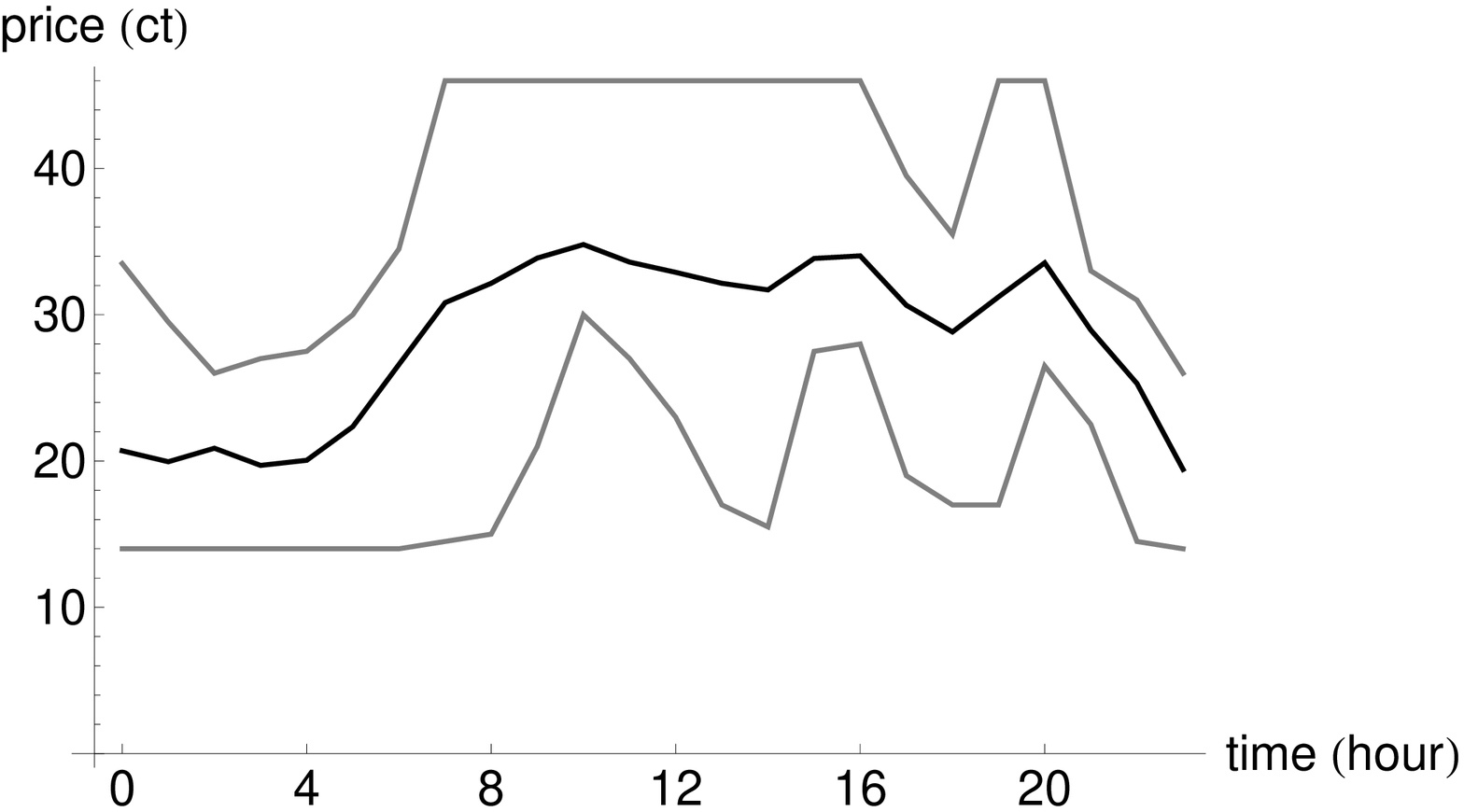}}
        \subfigure[February]{\label{fig:feb} \includegraphics[width = 0.23\textwidth]{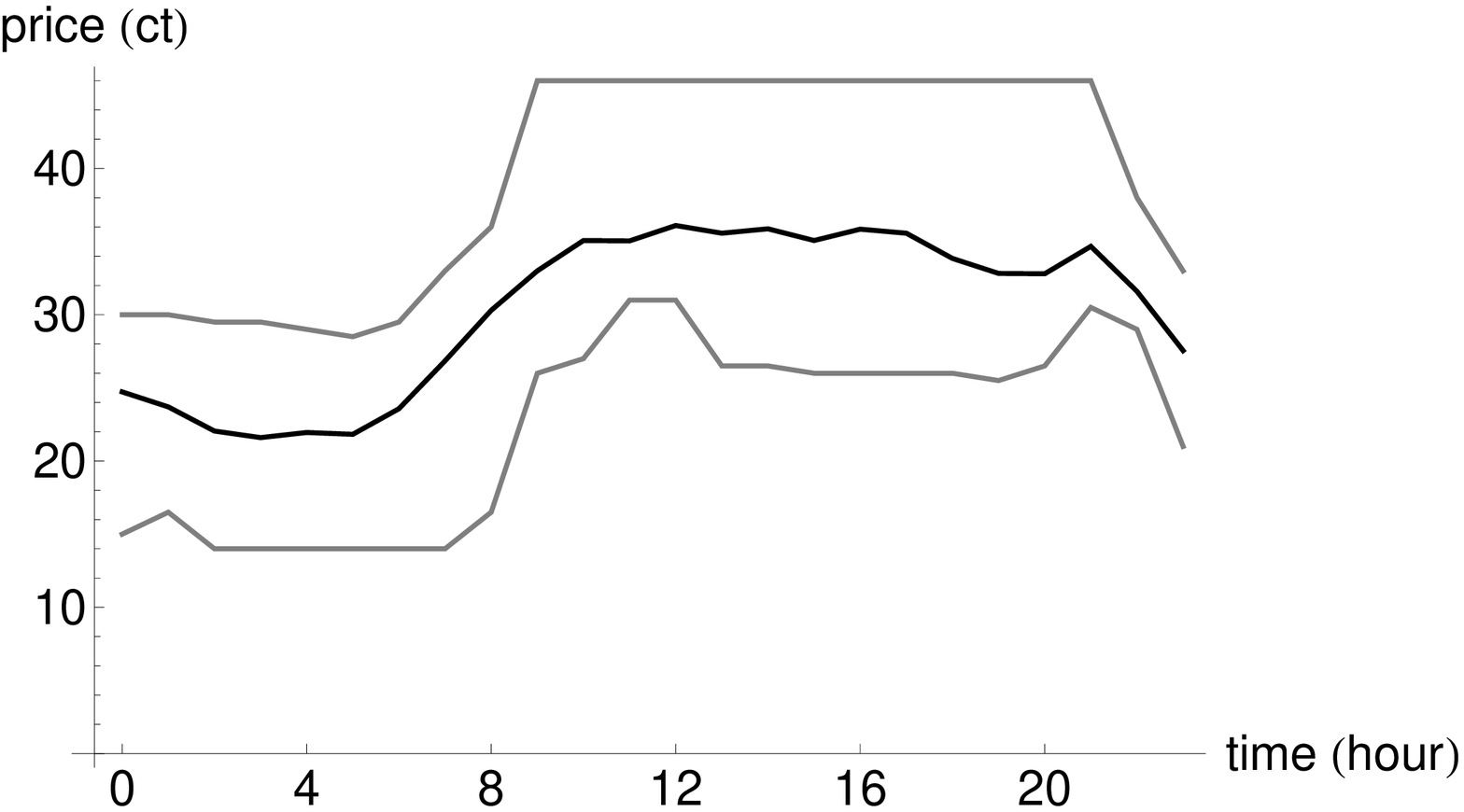}}\hspace{0.2cm}
    \end{center}
    \caption{Hourly average, minimum and maximum Ontario spot market prices (Ontario, 2011 dataset).}
    \label{fig:ontario_price}
\end{figure}

Figure~\ref{fig:jan} plots the average, minimum and maximum hourly price in Ontario during January 2011.
we observe one active phase with multiple price peaks between 9 a.m. and 10 p.m., and a low price at night.
Note that the energy prices are lower at night and display multiple peaks per day.
Figure~\ref{fig:feb} shows that the February prices follow a similar trend.

We first determine the empirical distribution of the January
2011 Ontario hourly energy prices and demand data. The empirical price (demand) distribution for each hour of the day
is computed by dividing the number of observations of a specific price (demand) level by the total number (31)
of observations. These distributions are used to determine the thresholds of the optimal policy.
Then, we use the February 2011 prices and demands to emulate the operation of the
optimal policy and compute the resulting electricity cost.

\subsection{Implementation}

Analytical determination of the optimal thresholds is possible for special cases but is difficult in general (see Section~\ref{sec:thresholds}). Instead, we compute the thresholds numerically using policy iteration.
Since policy iteration can be computationally intensive in practice, we have reduced the complexity of our implementation as follows.  First, we discretize the state space by rounding the demand data and energy storage level to multiples of 0.5~kWh, and the price data to multiples of 5~ct.  Second, we use the hour of the day as the modulating state and assume that prices and demand are independent from hour to hour (but not necessarily identically distributed). Thus, the optimal thresholds depend on price realization and the hour of the day, but are independent of the demand realization.
The modulating variable $\mathcal{M}$ could also be used to differentiate between different days and months and to take into account weather conditions. Such detailed state description would likely yield higher cost savings as it would allow more accurate price and demand predictions. On the other hand, it might generate too large a state space for policy iteration to be feasible. The following experiments demonstrate that using our simple choice of the modulating state yields very high cost savings.

We evaluate the optimal policy for a fully efficient battery ($\eta_d = \eta_c = 1$)
and no charging constraints ($\Amax_c(b) = \Amax_d(b) = \infty$) thus obtaining an upper bound on the
potential cost savings of energy storage.
%These numerical experiments can be easily extended to inefficient batteries.
We implement the policy iteration algorithm in Matlab; each slot in the policy iteration algorithm corresponds to one hour.
%and assume that slots last one hour.
For the above parameters and a discount factor $\alpha = 0.99$,
a laptop PC with a quad-core 2.2~GHz Intel processor and 8~GB RAM typically requires
approximately 5~min to compute the optimal thresholds.
%We use the January 2011 data to derive the price and demand empirical distributions and optimal thresholds
%and the February 2011 data for the numerical experiments themselves.

\subsection{Energy savings}\label{subsec:savings}

\begin{figure}[h]
    \begin{center}
        \includegraphics[width=1\linewidth]{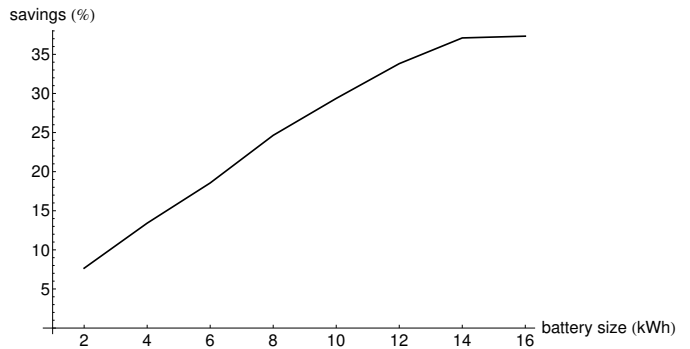}
    \end{center}
    \caption{Energy cost savings of optimal storage (over no energy storage) vs. battery size (Ontario, February 2011 dataset).}
    \label{fig:size}
\end{figure}

Figure~\ref{fig:size} shows the relative energy savings of the optimal policy over the setting without storage, as a function of the battery size $B_{\max}$. Three important observations are in place. First, the savings increase with battery size to up to 38\%, which is significant. Second, the savings reach their maximum at $B_{\max}$ = 16 (kWh); increasing battery size beyond this point does not increase savings, as the optimal policy will not utilize any battery capacity beyond 16~kWh. This saturation point can be explained by the fact that the value of stored energy decreases over time due to the discounting of the costs and the cyclic price and demand levels. Third, the size of a typical hybrid vehicle battery pack is in the order of 16~kWh~\cite{PWA10}. This suggests that car battery packs are well-suited for home energy storage since their size corresponds to the amount of storage required by the optimal energy storage policy.

%time-of-use pricing with three price levels. Time-of-use pricing is currently the most common implementation of dynamic pricing, and the price levels in this example correspond to the prices of the Ontario energy market~\cite{Ontario2010}.
%The demand reflects user profiles used by energy retailers to predict customer demand. We consider an efficient battery, i.e., $\eta_d = \eta_c = 1$, in order to obtain an upper bound on the cost savings in practice.

\subsection{Structure of the optimal policy}

\begin{figure}[h]
    \begin{center}
        \includegraphics[width=1\linewidth]{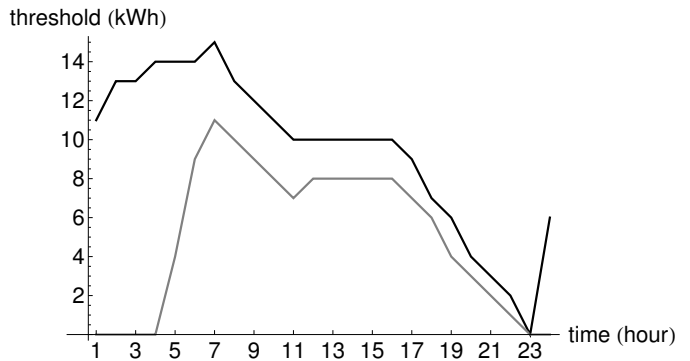}
    \end{center}
    \caption{Optimal thresholds vs hour of day, for 15~ct/kWh (black) and 25~ct/kWh (gray) price levels.}
    \label{fig:thresholds}
\end{figure}

In the remainder of this section we consider a battery with capacity of 16~kWh. To illustrate how the optimal storage policy works, we show in Figure~\ref{fig:thresholds} the thresholds as a function of time, for two price levels (15~ct and 25~ct). We observe that the thresholds peak early in the morning when the price is low. Moreover, the lower 15~ct price level yields lower thresholds at all hours, in accordance with Proposition~\ref{pro:monotone_thresholds}. Due to low demand and several consecutive hours with low prices, the 15~ct threshold drops to zero at 11 p.m., and increases again the next hour.

\begin{figure}[h]
    \begin{center}
        \includegraphics[width=1\linewidth]{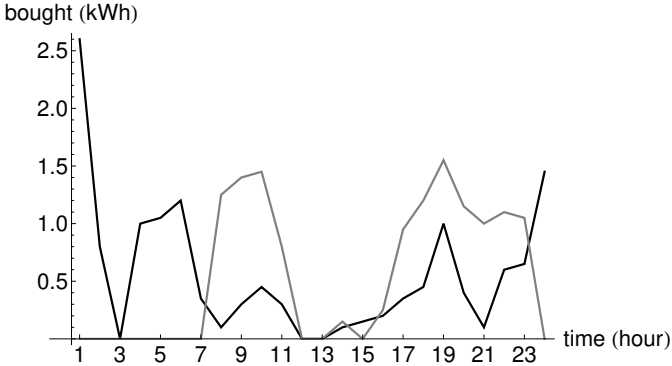}
    \end{center}
    \caption{The average amount of energy bought, with (black) and without (gray) storage.}
    \label{fig:buying_diff}
\end{figure}

Figure~\ref{fig:buying_diff} shows the average amount of energy bought, plotted against the hour of the day, with optimal energy storage (black line) and without storage (gray line). We observe that storage allows users to purchase energy early in the morning when the price is low, while users without storage are forced to buy energy during peak hours, at higher prices.

\subsection{Resource pooling}

\begin{figure}[h]
    \begin{center}
        \includegraphics[width=1\linewidth]{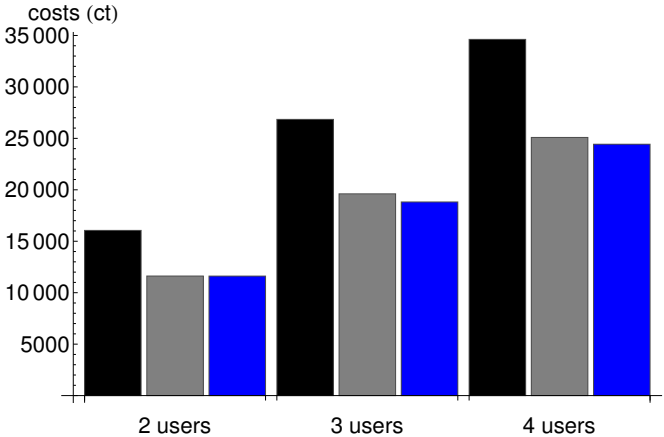}
    \end{center}
    \caption{The effect of resource pooling.}
    \label{fig:pooling}
\end{figure}

An alternative to energy storage for individual households is to pool storage capacity: rather than individual users each having a small battery, further cost savings might be achieved by multiple users sharing a single large battery. Figure~\ref{fig:pooling} compares the case of $n$ users each with a 16~kWh battery to the case of a shared $16 \times n$~kWh battery. We use the tool from~\cite{RTIC10} to generate distinct demand data for each individual home user. In the scenario with storage but without pooling, each home has its own storage, and each set of demand data corresponds to one energy storage unit. In this case, the thresholds are computed for each individual user. In the shared storage scenario, the aggregate demand data of all $n$ homes is input to the shared storage unit. Since this scenario only has a single storage unit (of size $16n$), we only have to compute the optimal thresholds once.
Figure~\ref{fig:pooling} compares the aggregate monthly costs without storage (black), storage without pooling (gray) and storage with pooling (blue) plotted against the number of users.

We observed similar performance by running the same experiment with artificially time-shifting the demands of these users to be negatively correlated (results not shown for brevity).
%We observe that pooling provides little benefits compared to individual storage.
This can be explained by the fact that the prices are the same for all users. Thus, irrespective of the size of the battery, %(and nature of demands if negative correlation graph is included)
the behavior of the optimal policy is primarily influenced by the common pricing signal, eliminating the potential benefits of pooling. %In fact, numerical experiments show that resource pooling provides only marginal savings, even if user demand is negatively correlated.

\section{Extensions \& Outlook}\label{sec:outlook}

Our model is a first step on energy storage management and focuses on a single user that utilizes storage to minimize its own costs.  The ideas and results can also be applied in multi-user settings where the objective is to minimize the grid peak load. For example, in a scenario with a large population, a certain fraction of battery-equipped users could apply the optimal control policy~\eqref{eqn:optimal_policy} to shift their demand from peak to off-peak hours.  Assessing the impact of such an approach is an interesting topic of future research.

The model can be extended in several directions to cover a broader variety of applications, and more realistic battery models. Below we briefly describe three such extensions.

%Throughout this paper we assumed that the battery does not need replacement, although in order to assess whether energy storage is profitable, battery replacement costs have to be taken into account as well. The battery lifetime is affected by the storage policy, and
%Battery lifetime benefits from longer sustained periods of charging and discharging.
%On one hand, this may complicate analysis because the optimal policy may depend on whether
%the battery was charged or discharged in previous slots. On the other hand, it may give rise
%to simple heuristics where the battery is alternatively fully charged and discharged.
%taking into account the costs for buying and replacing the battery introduces the problem
%of battery dimensioning. Smaller batteries are cheaper but may provide less opportunity
%to exploit price fluctuations. The optimal size of the battery will most likely depend on the spread and volatility of the energy prices, and may differ between energy markets, cf.~\cite{LF06}.
%Another tradeoff that comes up is between battery cost and efficiency, as cheap batteries are generally less efficient than their expensive counterparts.

\subsection{Battery replacement costs}
In order to assess the cost effectiveness of energy storage, one has to incorporate battery replacement costs. Determining the optimal policy in this case requires solving a joint problem of battery dimensioning and energy storage management for cost minimization or battery lifetime maximization.  Smaller batteries are cheaper but, as shown in Figure~\ref{fig:size}, they may provide less opportunity to exploit price fluctuations. The optimal size of the battery will most likely depends on the spread and volatility of energy prices and may differ between energy markets~\cite{LF06}.

As a first-order approximation of the battery replacement costs, we may assume that the battery breaks down after a geometric number of operations (charging/discharging), at which point the battery needs replacement at costs $C$. Under these assumptions, the analysis presented in this paper largely holds, with some minor modifications. Assuming that battery lifetime has mean $1/q$, the immediate costs can be rewritten as
\begin{equation*}
\gamma_{\bfx}(\delta) = (d(\bfx) + (\delta)^+ \eta_c^{-1} + (\delta)^- \eta_d) p(\bfx) + q C \indi{\delta \neq 0}.
\end{equation*}
This new cost function is convex everywhere but in $\delta = 0$, and Lemma~\ref{lem:cost_properties} and Theorem~\ref{thm:optimal} must be modified accordingly.

\subsection{Self-discharge and varying efficiency}

We may extend the battery model by including the effects of self-discharge and time-varying efficiency. That is, every time slot we assume that some fraction $\xi$ of the stored energy dissipates, and the storage evolution~\eqref{eqn:constr3} is modified accordingly:
\begin{equation*}
B(t+1) = \xi B(t) + \eta_c(t) A_2(t) - A_3(t).
\end{equation*}
The rest of the paper can be adjusted in a similar fashion, and all results and derivations continue to hold. The case $\xi = 1$ and $\eta_c(t) \equiv \eta_c$ corresponds to the present model.

\subsection{Local energy generation}

Another straightforward extension of our model is to assume that the end user itself generates some energy. This energy is either used to satisfy demand, is sold back to the grid or stored in the battery. This extension allows us to model for example wind energy farms, where energy storage can be used to successfully incorporate the increasing amount of renewable energy into the grid. Our results then extend to describe an optimal policy for the joint decision on buying, selling and storing (renewable) energy.

In order to make these adjustments, we must reinterpret $D(t)$ as the demand minus the amount of energy generated in a slot, and remove the constraints~\eqref{eqn:constr_pos} that state that $A_1(t)$, $A_2(t)$ and $A_3(t)$ are non-negative.

%Although batteries have seen a significant decrease in price and an increase in efficiency in recent years, it is not clear whether energy storage is economically viable as of now.

%It is essential to take into account battery replacement costs when assessing the overall effectiveness of energy storage. Although we have demonstrated that energy storage may lead to significant cost reduction for purchasing energy, these savings may be offset by the expensive storage system. Consequently, although batteries have seen a significant decrease in price and an increase in efficiency in recent years, it is not clear whether energy storage is economically viable as of now.

%\subsection{Energy market}
%
%We may ask ourselves what is the effect on the energy market when a significant fraction of users adopt energy storage. A possible outcome is that the resulting steady demand process will cause convergence of the energy market resulting in smaller price variations. While this is good from the perspective of both energy producers and users without energy storage, a less volatile price process will decrease the possibilities for exploiting price fluctuations for users with storage capacities. Consequently, the cost savings obtained from energy storage may decrease beyond the break-even point. The interplay between energy storage and the energy market is an interesting topic for future research.

\section{Conclusions}\label{sec:conclusions}

In this paper we studied the control of end-user energy storage under price fluctuations. We derived the structure of the cost-minimizing storage policy, which turns out to be a simple threshold-based policy~\eqref{eqn:optimal_policy}. %The thresholds only depend on the current energy price and are not affected by the buffer level.
We described the behavior of these thresholds for some special cases, and showed by means of a numerical study that energy storage can lead to significant cost savings. We discussed various model extensions and generalizations that would broaden the scope of this work, without fundamentally altering the results.

\appendix
\section{Remaining proofs}

\subsection{Proof of Lemma~\ref{lem:cost_properties}}\label{sec:proof_lemma}
\begin{IEEEproof}
Let $J_{\bfx,n}(b)$ denote the minimal $n$-step discounted costs, starting from state $\bfx$ and battery level $b$. Let $H_{\bfx,n}(\beta,b) =
\gamma_{\bfx}(\beta - b) + \alpha G_{\bfx,n}(\beta)$ and $G_{\bfx,n}(\beta) = \int\limits_{\bfy \in \Omega} f_{\bfx}(\bfy) J_{\bfy,n}(\beta){\rm d}\bfy$, then $J_{\bfx,n}(b)$ satisfies
\begin{equation}\label{eqn:bellman_finite}
J_{\bfx,n}(b) = \min_{\beta \in U_{\bfx}(b)} H_{\bfx,n-1}(\beta,b).
\end{equation}
These finite-horizon costs converge as $\lim_{n \rightarrow \infty} J_{\bfx,n} = J_{\bfx}$, cf.~\cite[Proposition 4.4]{BS96}. Thus, in order to show that $J_{\bfx}$ is convex and non-increasing, is suffices to show by induction that this holds for all $J_{\bfx,n}$.

This statement trivially holds for $n = 0$, since $J_{\bfx,0} \equiv 0$. Now let $n \in \mathds{N}$, and assume that $J_{\bfx,n-1}$ is convex and non-increasing. The operator on the right-hand side of~\eqref{eqn:bellman_finite} can be identified as the infimal convolution operator, which preserves convexity~\cite[Theorem 5.4]{Rockafellar72}. We have that $G_{\bfx,n-1}$ is convex by the induction hypothesis, and since it can be readily verified that $\gamma_{\bfx}$ is convex as well, so is $J_{\bfx,n}$.

Let $b_1, b_2 \in \Bu$, $b_1 \le b_2$, then in order to establish that $J_{\bfx,n}$ is non-increasing, we need to show that $J_{\bfx,n}(b_1) \ge J_{\bfx,n}(b_2)$. Denote by $\beta_{\bfx}^*$ the choice for $\beta$ he achieves the minimum in~\eqref{eqn:bellman_finite}. We distinguish two cases: (i) $\beta_{\bfx}^*(b_1) \in U_{\bfx}(b_2)$; and (ii) $\beta_{\bfx}^*(b_1) < U^-_{\bfx}(b_2)$. In case (i):
\begin{align*}
J_{\bfx,n}(b_2) &\le H_{\bfx,n-1}(\beta_{\bfx}^*(b_1),b_2)\\
&= \gamma_{\bfx}(\beta_{\bfx}^*(b_1) - b_2) + \alpha G_{\bfx,n-1}(\beta_{\bfx}^*(b_1))\\
&\le \gamma_{\bfx}(\beta_{\bfx}^*(b_1) - b_1) + \alpha G_{\bfx,n-1}(\beta_{\bfx}^*(b_1)) = J_{\bfx,n}(b_1),
\end{align*}
by the fact that $\gamma_{\bfx}$ is increasing.

In case~(ii) we have by the induction hypothesis
\begin{align*}
J_{\bfx,n}(b_2) &\le H_{\bfx,n-1}(U^-_{\bfx}(b_2),b_2)\\
&= \gamma_{\bfx}(-d(\bfx)) + H_{\bfx,n-1}(U_{\bfx}^-(b_2))\\
&\le \gamma_{\bfx}(\beta_{\bfx}^*(b_1) - b_1) + H_{\bfx,n-1}(\beta_{\bfx}^*(b_1)) = J_{\bfx,n}(b_1),
\end{align*}
completing the proof.
\end{IEEEproof}

\subsection{Proof of Theorem~\ref{thm:optimal}}\label{sec:proof_theorem}
\begin{IEEEproof}
Since $\gamma_{\bfx}$ and $J_{\bfx}$ are convex, so is $H_{\bfx}$, and~\eqref{eqn:bellman} can be recognized as a convex optimization problem. Thus, $\beta^* \in U_{\bfx}(b)$ is a global minimizer if and only if there exists some subgradient $g \in \partial H_{\bfx}(\beta^*,b)$ such that for all $\beta \in U_{\bfx}(b)$,
\begin{equation*}
g(\beta^* - \beta) \ge 0.
\end{equation*}
Thus, $\beta^* \in (\Uleft,\Uright)$ is a minimizer iff $0 \in \partial H_{\bfx}(\beta^*,b)$, while $\beta^* = \Uleft$ and $\beta^* = \Uright$ are minimizers iff there exists some $g \in \partial H_{\bfx}(\beta^*,b)$ such that $g \ge 0$ and $g \le 0$, respectively.

The subdifferentials of $\gamma_{\bfx}$ and $G_{\bfx}$ are given by
\begin{equation*}
\partial \gamma_{\bfx}(\delta) = \left\{
\begin{array}{ll}
\eta_d p(\bfx), & \delta < 0,\\
\lbrack\eta_d p(\bfx),\eta_c^{-1} p(\bfx)\rbrack, & \delta = 0,\\
\eta_c^{-1} p(\bfx), & \delta > 0,
\end{array}\right.
\end{equation*}
and
\begin{equation*}
\partial G_{\bfx}(\beta) = \lbrack\sleft(\beta),\sright(\beta)\rbrack,
\end{equation*}
for some $-\infty < \sleft(\beta) \le \sright(\beta) \le 0$. The subdifferential of $H_{\bfx}$ can then be written as (cf.~\cite[Theorem 23.8]{Rockafellar72}):
\begin{align*}
&\partial H_{\bfx}(\beta,b)\\
&= \left\{
\begin{array}{ll}
\lbrack\eta_d p(\bfx) + \alpha \sleft(\beta),\eta_d p(\bfx) + \alpha \sright(\beta)\rbrack, & \beta < b,\\
\lbrack\eta_d p(\bfx) + \alpha \sleft(\beta),\eta_c^{-1} p(\bfx) + \alpha \sright(\beta)\rbrack, & \beta = b,\\
\lbrack\eta_c^{-1} p(\bfx) + \alpha \sleft(\beta),\eta_c^{-1} p(\bfx) + \alpha \sright(\beta)\rbrack, & \beta > b.
\end{array}\right.
\end{align*}
Consequently, $\beta^* = \Uleft$ is optimal if
\begin{equation*}
\eta_d p(\bfx) + \alpha \sright(\beta^*) \ge 0.
\end{equation*}
Let $\beta^* \in (\Uleft,b)$, then $\beta^*$ is optimal if
\begin{equation*}
\eta_d p(\bfx) + \alpha \sleft(\beta^*) \le 0 \le \eta_d p(\bfx) + \alpha \sright(\beta^*).
\end{equation*}
Let $\beta^*  = b$, then $\beta^*$ is optimal if
\begin{equation*}
\eta_d p(\bfx) + \alpha \sleft(\beta^*) \le 0 \le \eta_c^{-1} p(\bfx) + \alpha \sright(\beta^*).
\end{equation*}
Let $\beta^* \in (b,\Uright)$, then $\beta^*$ is optimal if
\begin{equation*}
\eta_c^{-1} p(\bfx) + \alpha \sleft(\beta^*) \le 0 \le \eta_c^{-1} p(\bfx) + \alpha \sright(\beta^*).
\end{equation*}
Finally, let $\beta^*  = \Uright$, then $\beta^*$ is optimal if
\begin{equation*}
\eta_d p(\bfx) + \alpha \sleft(\beta^*) \le 0.
\end{equation*}

Denote
\begin{align}
\nonumber B_{1,\bfx}^- &= \{\beta \in [0,\Bmax]\ :\ \eta_c^{-1} p(\bfx) + \alpha \sright(\beta) \ge 0\},\\
\nonumber B_{2,\bfx}^- &= \{\beta \in [0,\Bmax]\ :\ \eta_c^{-1} p(\bfx) + \alpha \sleft(\beta) \le 0\},\\
\nonumber B_{1,\bfx}^+ &= \{\beta \in [0,\Bmax]\ :\ \eta_d p(\bfx) + \alpha \sright(\beta) \ge 0\},\\
B_{2,\bfx}^+ &= \{\beta \in [0,\Bmax]\ :\ \eta_d p(\bfx) + \alpha \sleft(\beta) \le 0\}, \label{eqn:sets}
\end{align}
and define
\begin{align*}
\beta_{1,\bfx}^- &= \left\{
\begin{array}{ll}
\min B_{1,\bfx}^-,   &   {\rm if~} B_{1,\bfx}^- \neq \emptyset,\\
\Bmax,   &   {\rm otherwise},
\end{array}
\right.
\\
\beta_{2,\bfx}^- &= \left\{
\begin{array}{ll}
\max B_{2,\bfx}^-,   &   {\rm if~} B_{2,\bfx}^- \neq \emptyset,\\
0,   &   {\rm otherwise},
\end{array}
\right.
\\
\beta_{1,\bfx}^+ &= \left\{
\begin{array}{ll}
\min B_{1,\bfx}^+,   &   {\rm if~} B_{1,\bfx}^+ \neq \emptyset,\\
\Bmax,   &   {\rm otherwise},
\end{array}
\right.
\\
\beta_{2,\bfx}^+ &= \left\{
\begin{array}{ll}
\max B_{2,\bfx}^+,   &   {\rm if~} B_{2,\bfx}^+ \neq \emptyset,\\
0,   &   {\rm otherwise}.
\end{array}
\right.
\end{align*}
Since $\sleft$ and $\sright$ are non-decreasing (due to the fact that $G_{\bfx}$ is convex and non-increasing), we have that $\beta_{1,\bfx}^- \le \beta_{2,\bfx}^- \le \beta_{1,\bfx}^+ \le \beta_{2,\bfx}^+$.

Then, for any $\beta_{\bfx}^- \in \lbrack \beta_{1,\bfx}^-,\beta_{2,\bfx}^-\rbrack$ and $\beta_{\bfx}^+ \in \lbrack \beta_{1,\bfx}^+,\beta_{2,\bfx}^+\rbrack$, the policy~\eqref{eqn:optimal_policy} is optimal.
\end{IEEEproof}

\begin{remark}
Note that the proof of Theorem~\ref{thm:optimal} describes a continuum of optimal policies, since any choice for $\beta_{\bfx}^- \in \lbrack \beta_{1,\bfx}^-,\beta_{2,\bfx}^-\rbrack$ and $\beta_{\bfx}^+ \in \lbrack \beta_{1,\bfx}^+,\beta_{2,\bfx}^+\rbrack$ defines a solution to~\eqref{eqn:min_problem}.
\end{remark}

\subsection{Proof of Corollary~\ref{col:efficient}}

\begin{IEEEproof}
In case $\eta_c = \eta_d = 1$, we see that $\lbrack \beta_{1,\bfx}^-,\beta_{2,\bfx}^-\rbrack = \lbrack \beta_{1,\bfx}^+,\beta_{2,\bfx}^+\rbrack$, and the result readily follows.
\end{IEEEproof}

\subsection{Proof of Proposition~\ref{pro:sufficient_conditions}}\label{sec:proof_sufficient}

\begin{IEEEproof}
We have that $\beta_{\bfx} = 0$ if for all $\beta \in U_{\bfx}(0)$,
\begin{align*}
&H_{\bfx}(0,0) \le H_{\bfx}(\beta,0)\\
\Leftrightarrow{}& d(\bfx) p(\bfx) - (d(\bfx) + \beta) p(\bfx)\\
&{}+ \alpha \int_{\bfy \in \Omega} f_{\bfx}(\bfy)\big( J_{\bfy}(0) - J_{\bfy}(\beta)\big) \le 0\\
\Leftrightarrow{}& \alpha \int_{\bfy \in \Omega} f_{\bfx}(\bfy)\big( J_{\bfy}(0) - J_{\bfy}(\beta)\big) \le \beta p(\bfx).
\end{align*}
This holds if $J_{\bfy}(b_1) - J_{\bfy}(b_2) \le (b_2 - b_1) \frac{p(\bfx)}{\alpha}$ for all $b_1 < b_2$, $\bfy \in \Omega$.

In order to verify that $\beta_{\bfx} = \Bmax$, we need to show that for all $\beta \in U_{\bfx}(\Bmax)$,
\begin{align*}
&H_{\bfx}(\Bmax,\Bmax) \le H_{\bfx}(\beta,\Bmax)\\
\Leftrightarrow{}& d(\bfx) p(\bfx) - (d(\bfx) + (\Bmax - \beta)) p(\bfx)\\
 &{}+ \alpha \int_{\bfy \in \Omega} f_{\bfx}(\bfy)\big( J_{\bfy}(\Bmax) - J_{\bfy}(\beta)\big) \le 0\\
\Leftrightarrow{}& \alpha \int_{\bfy \in \Omega} f_{\bfx}(\bfy)\big( J_{\bfy}(\beta) - J_{\bfy}(\Bmax)\big) \ge (\Bmax - \beta) p(\bfx).
\end{align*}
This holds if $J_{\bfy}(b_1) - J_{\bfy}(b_2) \ge (b_2 - b_1) \frac{p(\bfx)}{\alpha}$ for all $b_1 < b_2$, $\bfy \in \Omega$.
\end{IEEEproof}

\subsection{Proof of Proposition~\ref{pro:max_price}}\label{sec:proof_maxprice}

\begin{IEEEproof}
We will show that for all $b_1,b_2 \in \Bu$, $b_1 < b_2$ and $\bfx \in \Omega$,
\begin{equation}\label{eqn:proof_zer_thr1}
J_{\bfx}(b_1) - J_{\bfx}(b_2) \le (b_2 - b_1) \Pmax.
\end{equation}
It then readily follows that in both cases (i) and (ii), the first condition of Proposition~\ref{pro:sufficient_conditions} is satisfied.

In order to show that~\eqref{eqn:proof_zer_thr1} holds, we apply the finite horizon framework presented in the proof of Lemma~\ref{lem:cost_properties}, and use induction on the horizon $n$. First, it is readily seen that
\begin{equation*}
J_{\bfx,0}(b_1) - J_{\bfx,0}(b_2) = 0 \le (b_2 - b_1) \Pmax.
\end{equation*}
Now assume that
\begin{equation*}
J_{\bfx,n-1}(b_1) - J_{\bfx,n-1}(b_2) \le (b_2 - b_1) \Pmax.
\end{equation*}
Then
\begin{align*}
&J_{\bfx,n}(b_1) - J_{\bfx,n}(b_2)\\
={}& \big((\beta_{\bfx}^*(b_1) - b_1) - (\beta_{\bfx}^*(b_2) - b_2)\big) p(\bfx)\\
 &{}+ \alpha \int_{\bfy \in \Omega} f_{\bfx}(\bfy)\big( J_{\bfy,n-1}(\beta_{\bfx}^*(b_1)) - J_{\bfy,n-1}(\beta_{\bfx}^*(b_2)) \big){\rm d}\bfy\\
\le{}& \big((\beta_{\bfx}^*(b_1) - b_1) - (\beta_{\bfx}^*(b_2) - b_2)\big) \Pmax\\
 &{}+ \alpha(\beta_{\bfx}^*(b_2) - \beta_{\bfx}^*(b_1)) \Pmax  = (b_2 - b_1) \Pmax,
\end{align*}
completing the proof.
\end{IEEEproof}

\subsection{Proof of Proposition~\ref{pro:monotone_thresholds}}

\begin{IEEEproof}
For i.i.d.\ prices we see that the future prices do not depend on the current state, i.e., $G_{\bfx} \equiv G$. Consequently, we have that the subdifferential of $G$ is also independent of $\bfx$, and we write $\partial G(\beta) = [\sigma^-(\beta),\sigma^+(\beta)]$. Now, the sets $B_{1,\bfx}^-,B_{1,\bfx}^+,B_{2,\bfx}^-$ and $B_{2,\bfx}^+$ in~\eqref{eqn:sets} only depend on $\bfx$ through $p(\bfx)$, and are non-increasing in $p(\bfx)$ due to the non-decreasingness of $\sigma^-$ and $\sigma^+$. Thus, the thresholds are non-increasing in $p(\bfx)$ as well.
\end{IEEEproof}

\subsection{Details of Example~\ref{exa:example}}\label{sec:details_example}

We denote by $\bfx_i$ the state such that $p(\bfx_i) = i$, $i = 1,\dots,4$. First, observe that in this example the control set does not depend on the battery level, and $\beta_{\bfx}^*(b) \equiv \beta_{\bfx}^*$. Consequently, for all $\bfy \in \Omega$, $0 \le b_1 \le b_2 \le \Bmax$,
\begin{equation}
J_{\bfy}(b_1) - J_{\bfy}(b_2) = (b_2 - b_1)p(\bfy). \label{eqn:example1}
\end{equation}

By Proposition~\ref{pro:max_price} we know that $\beta_4 = 0$. In order to show that $\beta_1 = 1$, $\beta_2 = 0$ and $\beta_3 = 1$, we need to demonstrate that (cf. Proposition~\ref{pro:sufficient_conditions}):
\begin{align}
\nonumber &\frac{1}{2}(J_{\bfx_1}(\beta) - J_{\bfx_1}(1)) + \frac{1}{2}(J_{\bfx_3}(\beta) - J_{\bfx_3}(1))\\
\ge{}& p_1(1 - \beta) p_1/\alpha,\label{eqn:example2}\\
&J_{\bfx_1}(0) - J_{\bfx_1}(\beta) \le p_2 \beta /\alpha,\label{eqn:example3}\\
&J_{\bfx_4}(\beta) - J_{\bfx_4}(1) \ge p_3 (1 - \beta)/\alpha,\label{eqn:example4}
\end{align}
respectively.

By~\eqref{eqn:example1},
\begin{align*}
&\frac{1}{2}(J_{\bfx_1}(\beta) - J_{\bfx_1}(1)) + \frac{1}{2}(J_{\bfx_3}(\beta) - J_{\bfx_3}(1))\\
 ={}& 2(1 - \beta) \ge (1 - \beta)/\alpha,
\end{align*}
which corresponds to~\eqref{eqn:example2}. For $\beta_2$ we have
\begin{equation*}
J_{\bfx_1}(0) - J_{\bfx_1}(\beta) = \beta \le  \beta p_2/\alpha,
\end{equation*}
which satisfies~\eqref{eqn:example3}. Finally, we see that also~\eqref{eqn:example4} is satisfied since
\begin{equation*}
J_{\bfx_4}(\beta) - J_{\bfx_4}(1) = (1 - \beta) p_4 \ge (1 - \beta) p_3/\alpha.
\end{equation*}

\end{document}